# Simple and Onion-type Fullerenes shells as resonators and amplifiers

M. Ya. Amusia[1, 2]

[1]A. F. Ioffe Physical-Technical Institute, St.-Petersburg 194021, Russia
and
[2]Racah Institute of Physics, the Hebrew University, Jerusalem 91904, Israel

E-mail: Miron.Amusia@mail.ioffe.ru

**Abstract.** We discuss the influence of a single or double fullerenes shell upon photoionization and vacancy decay of an atom, stuffed inside the fullerenes construction. The main manifestations of this influence are additional structures in the photoionization cross-section and variation of the vacancy decay probability.

The main mechanisms, with which fullerenes shells affect the processes in caged atoms is the scattering of the outgoing electrons by the fullerenes shell and modification of the photon beam due to fullerenes shell polarization.

General consideration will be illustrated by numeric calculations where $C_{60}$ and $C_{240}$ will be chosen as fullerenes and Ar and Xe as caged atoms.

1. **Introduction**

Soon after discovery of fullerenes in 1985 it became clear that inside its empty shell an atom or even a small molecule can be "caged". Today inside a fullerene almost any element from the periodic table can be stuffed, thus forming new objects for scientific research and technological applications. An object that consist of an atom A caged inside a fullerenes shell constructed from N carbon atoms is called endohedral atom and is denoted as $A@C_N$.

Fullerenes and endohedrals are objects that are interesting as building blocks of new unusual materials and pure scientifically, as very specific nanostructures. Photoionization and photoelectron spectroscopy is an effective method to investigate structure and properties of these bodies. In this paper I will concentrate on endohedrals. It is not surprising that a great deal of attention has been and still is concentrated on their photoionization. From the point of view of photoionization studies they are similar to multi-shell heavy atoms.

Fortunately, the inner size of endohedrals is big enough to absorb an atom not destroying essentially its structure. On the other hand, the caged atom is not destroying the fullerene. So, in a sense, the inner atom A in $A@C_N$ serves as a "lamp" that shines "light" in the form of photoelectron waves that "illuminates" the fullerene $C_N$ from the inside. It is important that the atom inside, A, has different photoionization cross section and vacancy decay probabilities. So, by comparing photoelectron spectroscopy data of atom A and endohedral $A@C_N$, information could be obtained on the fullerene structure. The weaker is the A and $C_N$ coupling, the more reliable data are obtainable on the $C_N$ itself.

In principle, the "atom A - fullerene $C_N$" interaction leads to a number of consequences. Indeed, the atom A can be compressed by the $C_N$ or expanded. Respectively, $C_N$ can be expended or compressed. The photoelectrons from A in $A@C_N$ on their way out of the endohedral can collide with the fullerenes shell, both elastic and inelastic, thus emitting other electrons. The elastic scattering is a sort of reflection and refraction, leading to strong interference in the resulting outlook of the initial photoelectron wave.

The incoming beam of photons, while reaching the caged atom polarizes the fullerene electron cloud, induces in it multiple electrical and, in principal, magnetic moments. Since the fullerenes shell is big enough, the respective polarizabilities could be also big, so the photon beam that reaches the cage atom becomes considerably modified.

On top of it, the fullerene is a relatively big object, so one can expect unusual collective motion of there electrons and even different rather specific phase transitions, including of but not limited to finite system superconductivity and Fermi-condensation.

The vacancies initially created in the caged atom acquire a number of new pathways of decay as compared to the free atom, so the total decay picture can be modified even quantitatively.

We are very far from complete theoretical understanding of this picture and still there are quite a few experiments in this direction [1, 2]. In this situation it is natural to study at first a much more simplified picture. Assuming that we do not have specific collective effects in fullerene, like Fermi-condensation [3], the best theoretical approach to treat the phenomenon would be to apply to an endohedral, with its N carbon atom and the caged atom A, the theoretical approach known as the Random Phase Approximation with Exchange (RPAE) [4], using as the initial step the one-electron Hartree-Fock (HF) method.

In spite of tremendous achievements in computing, such a program seems to be still far from our real possibilities. So, further simplifications are required. As a first step it is assumed that at least even for medium fast outgoing electrons the inelastic processes in the fullerenes shell are negligible. Qualitative analyses of the existing experimental data for endohedral ionization by photons with energies from thresholds up to several hundreds electron-volts shows that one can substitute the real fullerenes structure by a layer of collectivized electrons that move in the field of homogeneously smeared positive charge of carbon nuclei screened each by two 1s-electrons. This simplification permits to deal with endohedrals photoionization reasonable accurate thus believing that the results obtained are at least qualitatively reliable.

We assume further that the fullerenes shell can be substituted by a rather simple potential, like a square step or even a zero-thickness delta-function, spherically-symmetric for such fullerenes as $C_{60}$ or $C_{240}$ potential. The atom A, located well inside $A@C_N$ can be easily treated in HF and RPAE, while the "fullerene – atom A" interaction can be considered as a sort of perturbation. Additional calculation simplicity is achieved by using as A objects, the nucleus of which is located at the center of the fullerenes shell.

In the frame of these simplifying assumptions a number of theoretical studies were performed and they determine the direction of concentration of current efforts in this field. It was demonstrated in a number of papers (e.g. [5 -7] and reference therein), in the frame of the outlined general approach that the $C_{60}$ shell adds prominent resonance structure in the photoionization cross section of endohedral atoms. Almost nothing is known about the effect of $C_{240}$. The combined action of $C_{60}$ and $C_{240}$ in onion fullerene that can be denoted as $A@C_{60}@C_{240}$ is considered literally in only a couple of papers [8, 9].

In this paper we will consider the complicated resonance structure of the endohedral atom photoionization cross section. Namely, very strong enhancement and interference patterns in the photoionization cross-section of the valent and subvalent subshells of noble gas endohedral atoms $A@C_{60}$ and $A@C_{60}@C_{240}$ are demonstrated. It is shown also that the atomic Giant resonance can be either completely destroyed or remains almost untouched depending on the velocity of photoelectrons that are emitted in the resonance's decay process.

It exist a deep similarity between a multi-electron atom and an endohedral from the point of view of the multi-electron effects. Using the presented above simplified description of the electron fullerenes shell, an endohedral can be treated as an atom with an additional multi-electron shell. In the same way one can consider multi-layer fullerenes or so-called fullerenes onions, in which in the central empty volume the caged atom is located.

The pronounced action of the multi-electron neighboring shell upon a few-electron one was considered for the first time thirty-five years ago. As an example, the influence of $3p^6$ electrons upon the $3s^2$ in Ar has been presented [10]. A more complicated case with three interacting subshells was considered in [11]. There it was demonstrated that the $5p^6$ and $4d^{10}$ subshells act upon $5s^2$ in Xe very strong, completely modifying the $5s^2$ photoionization cross-section.

In endohedrals as compared to free atoms A an essential simplifying factor exists. Namely, the radius of the fullerene shell R significantly exceeds that of an encapsulated atom $r_A$. This makes it possible for photoionization of the A atom in the first approximation to consider the electronic sub-systems of the fullerene shell (or several shells in fullerene onions) and atom as practically independent of each other. For this reason, the amplitude of atom photoionization going through virtual excitation of fullerene shell electrons can be expressed directly via the dynamic polarizability of the fullerene shell $\alpha^d_{C_{60}}(\omega)$, or in case of "onions", of $\alpha^d_{C_{60}}(\omega)$ and $\alpha^d_{C_{240}}(\omega)$. In those cases when the frequency of electromagnetic radiation is close to frequencies of plasma oscillations of the collectivized electrons of the fullerene, the role of this two-step process becomes decisively important, as the role of $4d^{10}$ upon $5s^2$ in isolated Xe.

Along with fullerenes shell polarization, one has to take into account also the reflection and refraction of the photoelectron wave, which goes from $np^6$ and $ns^2$ subvalent shell, by the static potential of the fullerene. This reflection leads to formation of oscillating pattern of the cross section (see e.g. [12, 13]).

The molecule $A@C_N$ is a remarkable concrete example illustrating the role of the inter-shell interactions in the fullerene-like molecules, qualitatively similar but even much stronger than in the isolated atoms. In this case three types of resonances, namely that in photoelectrons' reflection from the fullerene shell, effect of plasmon-type excitation of generalized $C_{60}$ electrons upon the photoionization of atom's A outer shells and that caused by action of caged atoms neighboring to $np$ and $ns$-subshells, interfere.

We will discuss here three types of manifestation of the $C_{60}$ effect upon the endohedral atom photoionization. We will consider the action of the $C_{60}$ polarization by the incoming photon beam and photoelectron's reflection by the $C_{60}$ static potential upon photoionization of outer $np^6$ subshell of noble gases. We will show that the combination of both effects leads to increase of the cross-sections by a factor 40-45, creating powerful resonances that we call *Giant endohedral resonances*.

The effect of $C_{60}$ polarization on $ns^2$ is much less pronounced then on $np^6$ but still, together with static potential reflection adds very interesting structure that include also the effect of the neighboring $np^6$ subshell. A simple method was developed to take into account the reflection of photoelectron wave by $C_{60}$ and $C_{240}$ static potential that was presented by a zero-thickness $\delta$-type so-called bubble potential (see [14] and references there in). This approximation is valid for slow photoelectrons, whose wavelength is bigger than the thickness of $C_{60}$ and $C_{240}$ shells.

We will consider the role $C_{60}$ upon the atomic Giant resonances. Since their energy is high enough, the role of $C_{60}$ polarization is negligible. Starting with Xe@$C_{60}$, where the Giant resonance is almost completely destroyed by the photoelectron's reflection, we will end up with Eu@$C_{60}$, where the atomic Giant resonance remains almost untouched.

We will show how reflection of Auger electrons and fullerenes shell polarization affects the vacancy decay in the caged atom.

Entirely, we will show in this paper that the dynamic polarization of $C_N$ drastically modifies the outer [7, 15] and subvalent [16] shell photoionization cross section at any frequency of the incoming radiation $\omega$. The photoionization cross-section of outer and subvalent shells of noble gas endohedral atoms essentially differ from respective data for isolated atoms. We will also trace the modification of the atomic Giant resonances under the $C_{60}$ shell action on the way from Xe@$C_{60}$ to Eu@ $C_{60}$ [17].

## 2. Essential formulae

We will use here theoretical approaches already developed in a number of previous papers [12, 13]. However, for completeness, let us repeat the main points of the consideration and present the essential formula used in calculations.

Let us start with the problem of an isolated spherical endohedral A@$C_N$ or A@$C_{N1}$@$C_{N2}$ denoted in formulas as AC.

The differential in angle $d\Omega$ cross-section of the photoelectron's emission under the action of non-polarized light is given for spherically-symmetric endohedral shell by the following expression [18]

$$\frac{d\sigma_{nl}^{AC}(\omega)}{d\Omega} = \frac{\sigma_{nl}^{AC}(\omega)}{4\pi}[1 - \frac{1}{2}\beta_{nl}^{AC}(\omega)P_2(\cos\theta) + \kappa\gamma_{nl}^{AC}(\omega)P_1(\cos\theta) + \kappa\eta_{nl}^{AC}(\omega)P_3(\cos\theta)]. \quad (1)$$

Here $\sigma_{nl}^{AC}(\omega)$ is the $nl$-subshell partial cross-section, $\kappa = \omega/c$, $c$ is the speed of light and $P_i(\cos\theta)$ are the Legendre polynomials, $\theta$ is the angle between directions of incoming light beam and velocity of the outgoing electron, $\beta_{nl}^{AC}(\omega)$ is the dipole, while $\gamma_{nl}^{AC}(\omega)$ and $\delta_{nl}^{AC}(\omega)$ are so-called non-dipole angular anisotropy parameters.

There are two possible dipole transitions from subshell $l$, namely $l \to l \pm 1$ and three quadrupole transitions $l \to l; l \pm 2$.

The cross-section $\sigma_{nl}^{AC}(\omega)$ is determined by the dipole $D_{nl \to \varepsilon l \pm 1}^{AC}(\omega)$ amplitude, that for simplicity of relations we will denote as $D_{l \pm 1}^{AC} \equiv D_{nl \to \varepsilon l \pm 1}^{AC}(\omega)$ that leads to the following expression[1]

$$\sigma_{nl}^{AC}(\omega) = \frac{16\pi^2\omega}{3c}\left[(l+1)\left|D_{l+1}^{AC}\right|^2 + l\left|D_{l-1}^{AC}\right|^2\right] \quad (2)$$

Corresponding general expressions for $\beta_{nl}^{AC}(\omega)$, $\gamma_{nl}^{AC}(\omega)$ and $\eta_{nl}^{AC}(\omega)$ are formed by dipole and combinations of dipole $D_{l \pm 1}^{AC}(\omega)$ and quadrupole $Q_{nl \to \varepsilon l; l \pm 2}^{AC}(\omega) \equiv Q_{l;l \pm 2}^{AC}$ matrix elements of photoelectron transitions and photoelectrons waves scattering phases $\delta_{l'}$. It is these expressions that are employed in calculations. Using [4] we have:

---

[1] Atomic system of units is used in this paper with electron charge $e$, mass $m$, and Planck constant $\hbar$ equal to 1.

$$\beta_{nl}^{AC}(\omega) = \frac{1}{(2l+1)\left[(l+1)\left(\tilde{D}_{l+1}^{AC}\right)^2 + l\left(\tilde{D}_{l-1}^{AC}\right)^2\right]}\left[(l+1)(l+2)\left(\tilde{D}_{l+1}^{AC}\right)^2 + l(l-1)\left(\tilde{D}_{l-1}^{AC}\right)^2 - 6l(l+1)\left(\tilde{D}_{l+1}^{AC}\right)\left(\tilde{D}_{l-1}^{AC}\right)\cos(\tilde{\delta}_{l+1} - \tilde{\delta}_{l-1})\right]. \quad (3)$$

Parameters $\gamma_{nl}^{AC}(\omega)$ and $\eta_{nl}^{AC}(\omega)$ are given by the following expression [19]

$$\gamma_{nl}^{AC}(\omega) = \frac{3}{5\left[(l+1)\tilde{D}_{l+1}^2 + l\tilde{D}_{l-1}^2\right]} \times$$
$$\times \left\{\frac{(l+1)}{2l+3}[3(l+2)\tilde{Q}_{l+2}\tilde{D}_{l+1}\cos(\tilde{\delta}_{l+2} - \tilde{\delta}_{l+1}) - l\tilde{Q}_l\tilde{D}_{l+1}\cos(\tilde{\delta}_{l+2} - \tilde{\delta}_{l+1})] - \right. \quad (4)$$
$$\left. - \frac{l}{2l+1}\left[3(l-1)\tilde{Q}_{l-2}\tilde{D}_{l-1}\cos(\tilde{\delta}_{l-2} - \tilde{\delta}_{l-1}) - (l+1)\tilde{Q}_l\tilde{D}_{l-1}\cos(\tilde{\delta}_l - \tilde{\delta}_{l-1})\right]\right\},$$

$$\eta_{nl}^{AC}(\omega) = \frac{3}{5\left[(l+1)\tilde{D}_{l+1}^{AC} + l\tilde{D}_{l-1}^{AC}\right]} \left\{\frac{(l+1)(l+2)}{(2l+1)(2l+3)}\tilde{Q}_{l+2}^{AC}\left[5l\tilde{D}_{l-1}^{AC}\cos(\tilde{\delta}_{l+2} - \tilde{\delta}_{l-1}) - \right.\right.$$
$$\left. - (l+3)\tilde{D}_{l+1}^{AC}\cos(\tilde{\delta}_{l+2} - \tilde{\delta}_{l-1})\right] - \frac{(l-1)l}{(2l-1)(2l+1)}\tilde{Q}_{l-2}^{AC} \times$$
$$\times \left[5(l+1)\tilde{D}_{l+1}^{AC}\cos(\tilde{\delta}_{l-2} - \tilde{\delta}_{l+1}) - (l-2)\tilde{D}_{l-1}^{AC}\cos(\tilde{\delta}_{l-2} - \tilde{\delta}_{l-1})\right] + \quad (5)$$
$$\left. + 2\frac{l(l+1)}{(2l-1)(2l+3)}\tilde{Q}_l^{AC}\left[(l+2)\tilde{D}_{l+1}^{AC}\cos(\tilde{\delta}_l - \tilde{\delta}_{l+1}) - (l-1)\tilde{D}_{l-1}^{AC}\cos(\tilde{\delta}_l - \tilde{\delta}_{l-1})\right]\right\}.$$

Here the following notations are used for dipole $D_{l\pm1}^{AC}(\omega) = \tilde{D}_{l\pm1}^{AC}(\omega)\exp(i\Delta_{l\pm1})$ and quadrupole $Q_{l;l\pm2}^{AC}(\omega) = \tilde{Q}_{l;l\pm2}^{AC}(\omega)\exp(i\Delta_{l,l\pm2})$ matrix elements and $\tilde{\delta}_{l'} \equiv \delta_{l'} + \Delta_{l'}$ with $\delta_{l'}$ denoting the outgoing electron scattering phase.

Very often experimentalists are using non-dipole parameters $\gamma_{nl}^{AC(c)}$ and $\delta_{nl}^{AC(c)}$, introduced in [20]. The following formula connect them to those defined by (4) and (5)

$$\gamma_{nl}^{AC(c)}/5 + \delta_{nl}^{AC(c)} = \kappa\gamma_{nl}^{AC}, \qquad \gamma_{nl}^{AC(c)}/5 = -\kappa\eta_{nl}^{AC}. \quad (6)$$

For any multi-particle system the RPAE equations for the dipole matrix elements looks like [4, 21]

$$\langle v_2|D(\omega)|v_1\rangle = \langle v_2|\hat{d}|v_1\rangle + \sum_{v_3,v_4}\frac{\langle v_3|D(\omega)|v_4\rangle(n_{v_4} - n_{v_3})\langle v_4 v_2|U|v_3 v_1\rangle}{\varepsilon_{v_4} - \varepsilon_{v_3} + \omega + i\eta(1 - 2n_{v_3})}, \quad (7)$$

where

$$\langle v_1 v_2|U|v_1' v_2'\rangle \equiv \langle v_1 v_2|V|v_1' v_2'\rangle - \langle v_1 v_2|V|v_2' v_1'\rangle. \quad (8)$$

Here $V \equiv 1/|\vec{r} - \vec{r}'|$ and $v_i$ is the total set of quantum numbers that characterize a HF one-electron state on discrete (continuum) levels, $\varepsilon_{v_i}$ is the HF energy, $\eta \to +0$. The set $v$ includes the principal quantum number (energy), angular momentum, its projection and the projection of the electron spin. The function $n_{v_i}$ (the so-called step-function) is equal to 1 for occupied and 0 for vacant one-electron states. Note that $\langle v_2|d|v_1\rangle$ is the dipole matrix element in the HF approximation with $\hat{d} = rP_1(\cos\theta)$

The matrix elements $D_{l\pm 1}$ are obtained by solving the radial part of the RPAE equation (7). As to the quadrupole matrix elements $Q_{l\pm 2,0}$, they are obtained by solving the radial part of the RPAE equation, similar to (7)

$$\langle v_2|Q(\omega)|v_1\rangle = \langle v_2|\hat{q}|v_1\rangle + \sum_{v_3,v_4} \frac{\langle v_3|Q(\omega)|v_4\rangle(n_{v_4} - n_{v_3})\langle v_4 v_2|U|v_3 v_1\rangle}{\varepsilon_{v_4} - \varepsilon_{v_3} + \omega + i\eta(1 - 2n_{v_3})}. \qquad (9)$$

Here in $r$-form one has $\hat{q} = r^2 P_2(\cos\theta)$.

Equations (7, 9) for finite and non-homogeneous objects have to be solved numerically. This is achieved for atoms, and corresponding procedures are discussed at length in [21]. Similar procedure of a "brutal force" approach could be applied in principle to endohedral atoms. However in reality this is unachievable on the current and foreseeable future computing level. Moreover, "brutal force" computation hides the picture of acting essential mechanisms and their relative role. This is particularly important for endohedral where essential simplifications of (7) and (9) can be achieved by using the already mentioned above facts: the fullerenes radius R is much bigger than the atomic radius $r_A$, $R \gg r_A$ and fullerenes shell thickness $\Lambda$ is much smaller than R, $\Lambda \ll R$. These features permits to neglect the real atom-like structure of $C_N$ substituting the shell by a bubble-type pseudo potential and taking into account semi- phenomenological the dynamic reaction of the fullerenes shell to the incoming photon beam.

## 3. Static effect of the fullerene shell

Let us start with the confinement effects. These effects near the photoionization threshold can be described within the framework of the bubble potential model. According to it, for small photoelectron energies the real static and not perfectly spherical potential of the $C_N$ can be presented by the zero-thickness pseudo-potential (see [22] and references therein):

$$V(r) = -V_0 \delta(r - R). \qquad (10)$$

The parameter $V_0$ is determined by the requirement that the binding energy of the extra electron in the negative ion $C_N^-$ is equal to its observable value. Addition of (10) to the atomic HF the potential leads to *reflection amplitude factor* $F_{l'}(k)$ in the continuous spectrum electron wave function and hence in the photoionization amplitudes, which depends only upon the photoelectron's linear momentum $k$ and orbital quantum number $l'$ [22]:

$$F_{l'}(k) = \frac{k \sin \Delta \delta_{l'}^R(k)}{2V_0 u_{\varepsilon l'}^2(R)},$$

$$\tan \Delta \delta_{l'}^R(k) = \frac{u_{\varepsilon l'}^2(R)}{u_{\varepsilon l'}(R)v_{\varepsilon l'}(R) + k/2V_0}, \quad (11)$$

where $\Delta \delta_{l'}^R(k) \equiv \Delta \delta_{l'}^R$ are the additional phase shifts due to the reflection by the fullerene shell potential (10). In these formulas $u_{\varepsilon l'}(r)$ and $v_{\varepsilon l'}(r)$ are the regular and irregular solutions of the atomic HF equations for a photoelectron with momentum $k = \sqrt{2\varepsilon}$, where $\varepsilon$ is the photoelectron energy, connected with the photon energy $\omega$ by the relation $\varepsilon = \omega - I_A$ with $I_A$ being the atom A ionization potential.

In a two-shell endohedral $A@C_{N1}@C_{N2}$ instead of (10) the following potential reproduces the static action of the fullerenes shells

$$V(r) = -V_1 \delta(r - R_1) - V_2 \delta(r - R_2), \quad (12)$$

where $R_2 > R_1$ are the radiuses of the two shells.

Addition of (12) leads to the following expressions for $F_l(k)$ and $\Delta \delta_l^R$ [9]:

$$F_{l'} \equiv F_{l'}(k) = \frac{k \sin \Delta \delta_{l'}^R}{2\left[V_1 u_1^2 + V_2 u_2^2 - 2V_1 V_2 u_1 u_2 (u_1 v_2 - u_2 v_1)/k\right]}. \quad (13)$$

$$\tan \Delta \delta_{l'}^R(k) = \frac{u_1^2 + u_2 V_2 \left[u_2/V_1 + 2u_1 V_1 (u_2 v_1 - u_1 v_2)/k\right]}{u_1 v_1 + k/2V_1 + u_2 v_2 V_2/V_1 - 2u_1 v_2 V_2 (u_2 v_1 - u_1 v_2)/k}. \quad (14)$$

Here $u_{1,2} \equiv u_{\varepsilon l'}(R_{1,2})$ and $v_{1,2} \equiv v_{\varepsilon l'}(R_{1,2})$.

Note that already from (13) a qualitative possibility of enhancement of $F_{l'}(k)$ due to two-shell structure of the atom A surrounding is seen: the denominator includes sign variation terms and therefore can be decreased.

Using the expressions for $F_{l'}(k)$, one can obtain the following relation for $D^{AC(r)}$ and $Q^{AC(r)}$ amplitudes for endohedral atom with account of photoelectron's reflection and refraction by the static potential (10) or (12), expressed via the respective values for isolated atom that correspond to $nl \to \varepsilon l'$ transitions $D_{l\pm 1}^A$ and $Q_{l,l\pm 2}^A$, respectively:

$$D_{l\pm 1}^{AC(r)}(\omega) = F_{l\pm 1}(k) D_{l\pm 1}^A(\omega),$$
$$Q_{l,l\pm 2}^{AC(r)}(\omega) = F_{l,l\pm 2}(k) Q_{l,l\pm 2}^A(\omega). \quad (15)$$

Assuming that the refracting action of the fullerene shell is weak, the following relation for the cross-section can be obtained

$$\sigma_{l\pm 1}^{AC(r)}(\omega) = [F_{l\pm 1}(k)]^2 \sigma_{l\pm 1}^A(\omega), \quad (16)$$

where $\sigma^A_{l\pm1}(\omega)$ is the contribution of the $nl \to \varepsilon l \pm 1$ transition to the photoionization cross-section of atomic subshell $nl$, $\sigma^A_{nl}(\omega)$. The relation (16) is valid for even experimental cross-section. Due to reflection the total scattering phase shift is determined by relation $\delta_{l'} = \delta^A_{l'} + \Delta^R_{l'}$.

If the action of the fullerene shell is not weak, it modifies not only the final outgoing electron amplitude and phase, but also the virtual continuous spectrum states $\nu_3(\nu_4)$ that is taken into account by corresponding factors $F_{l'}(k) \equiv F_{l'}$. In the frame of RPAE this is achieved by following substitution in (7) and (9): $(n_{\nu_4} - n_{\nu_3}) \to F^2_{\nu_3}n_{\nu_4}(1-n_{\nu_4}) - F^2_{\nu_4}n_{\nu_3}(1-n_{\nu_4})$.

Solutions of corrected in such a way equations (7) and (9) are denoted as $D_{l\pm1}(\omega)$ and $Q_{l,l\pm2}(\omega)$ instead of $D^A_{l\pm1}(\omega)$ and $Q^A_{l,l\pm2}(\omega)$, respectively.

## 4. Dynamic effect of the fullerene shell

Now let us discuss the role of polarization of the fullerene shell under the action of the photon beam [22]. The effect of the fullerene electron shell polarization upon atomic photoionization amplitude can be taken into account in RPAE using (7) and (9). This approximation is good for isolated atoms [4] and it is reasonable to assume that it is good also for endohedral atoms as well.

Taking into account the relation $R_1 \gg r_A$, one can neglect the exchange between atomic and fullerenes electrons, and present the interaction $U$ in (7) and (9) as a product For the dipole component it is

$$U_{CA} \approx \mathbf{r}_C \cdot \mathbf{r}_A / r^3_C, \quad (r_C \gg r_A). \quad (17)$$

where $\mathbf{r}_A$ and $\mathbf{r}_C$ are the atomic and fullerenes radii, respectively.

Using (17) the formula (9) is simplified considerably and is transformed from integral to algebraic equation that can be easily solved leading to [22]:

$$D^{AC(p)}_{l'}(\omega) \cong D^A_{l'}\left(1 - \frac{\alpha^d_C(\omega)}{R^3}\right) \equiv G^d_C(\omega)D^A_{l'}, \quad (18)$$

where $\alpha^d_C(\omega)$ is the dipole dynamical polarizability of the fullerene. The function $G^d_C(\omega)$ is called *polarization amplitude factor* and describes the modification of the caged atom photoionization amplitude due to fullerene shell dipole polarization under the action of the incoming photon beam.

Combining (18) and (15) we take into account the action of fullerenes shell polarization and photoelectron scattering by the fullerenes potential that leads to the following relation for the amplitude of endohedral atom's photoionization due to $nl \to \varepsilon l'$ transition $D^{AC}_{nl,\varepsilon l'}(\omega)$ with all essential atomic correlations taken into account [22, 9]:

$$D^{AC}_{nl,\varepsilon l'}(\omega) \cong F_{l'}(k)\left(1 - \frac{\alpha^d_C(\omega)}{R^3_C}\right)D^A_{nl,\varepsilon l'}(\omega) \equiv F_{l'}(k)G^d_C(\omega)D^A_{nl,\varepsilon l'}(\omega), \quad (19)$$

For the quadrupole amplitude one obtains, starting from (9), a similar expression:

$$Q_{nl,\varepsilon l'}^{AC}(\omega) \cong F_{l'}(k)\left(1 - \frac{\alpha_C^q(\omega)}{4R_C^5}\right)Q_{nl,\varepsilon l'}^A(\omega) \equiv F_{l'}(k)G_C^q(\omega)Q_{nl,\varepsilon l'}^A(\omega), \tag{20}$$

where $\alpha_C^q(\omega)$ is the quadrupole dynamical polarizability of the fullerene. The $G^{d,q}(\omega)$ factors are complex numbers that we present as

$$G^{d,q}(\omega) = \tilde{G}^{d,q}(\omega)\exp[i\eta^{d,q}(\omega)], \tag{21}$$

where $\tilde{G}^{d,q}(\omega)$ are respective absolute values.

According to (2), the cross-section acquires a factor $S(\omega) = [\tilde{G}^d(\omega)]^2$ that can be called *radiation enhancement parameter*.

The presence of a second deformable fullerene shell with the radius $R_2$ makes the situation considerably more complex. To simplify derivation of respective expressions we assume, on the ground of known real situation that $R_2 \gg R_1$ and can use the ratio $R_1/R_2 \ll 1$ as a small parameter. In this case the following expressions for $G_{C_{12}}^q(\omega)$ and $G_C^q(\omega)$ were obtained [9]:

$$G_{12}^d(\omega) \approx \left[1 - \left(\frac{\alpha_1}{R_1^3} + \frac{\alpha_2}{R_2^3}\right)\frac{1 - \frac{\alpha_1\alpha_2}{\alpha_1 R_2^3 + \alpha_2 R_1^3}\left(1 + \frac{R_1^3}{R_2^3}\right)}{1 - \frac{\alpha_1\alpha_2}{R_2^6}}\right], \tag{22}$$

where $\alpha_1$, $\alpha_2$ are dipole dynamic polarizabilities of shells 1 and 2 and

$$G_{12}^q(\omega) \cong \left[1 - \left(\frac{\alpha_1^q}{4R_1^5} + \frac{\alpha_2^q}{4R_2^5}\right)\frac{1 - \frac{\alpha_1^q\alpha_2^q}{4\alpha_1^q R_2^5 + 4\alpha_2^q R_1^5}\left(1 + \frac{R_1^5}{R_2^5}\right)}{1 - \frac{\alpha_1^q\alpha_2^q}{16R_2^6}}\right], \tag{23}$$

where $\alpha_1^q$, $\alpha_2^q$ are quadrupole dynamic polarizabilities of shells 1 and 2.

It is essential that the interaction between fullerenes shells is taken in (22) and (23) within RPAE frame, i.e. non-perturbatively.

As to the amplitudes $D_{nl,\varepsilon l'}^{AC}(\omega)$ and $Q_{nl,\varepsilon l'}^{AC}(\omega)$, they are determined for two-shell fullerenes by (19) and (20) with $G_{12}^d(\omega)$ and $G_{12}^q(\omega)$ instead of $G_C^d(\omega)$ and $G_C^q(\omega)$. To account for strong reflection, one has to substitute in (19) and (20) also $D_{nl,\varepsilon l'}^A(\omega)$ and $Q_{nl,\varepsilon l'}^A(\omega)$ by $D_{nl,\varepsilon l'}(\omega)$ and $Q_{nl,\varepsilon l'}(\omega)$ (see definition at the end of Section 3).

All additional structure in characteristics of photoionization of endohedrals as compared to isolated atoms comes from the interference oscillations of $F_{l'}(k)$ as a function of $k$ and resonance behavior of $G_C^{d,q}(\omega)$ as a function of $\omega$ and additional variation of the cross section due to effect of $F_{l'}(k)$ upon intermediate atomic states in (7) and (9).

## 5. Vacancy decay

The presence of fullerenes shells modifies not only the endohedral photoionization characteristics but affects also the vacancy decay probability of the caged atom. There is a whole variety of modifications, among which we concentrate on two: the scattering of an outgoing Auger electron upon the fullerenes shell static potential described approximately by (10) or (12) and absorption by the fullerenes shell of the virtual or real photon emitted while the vacancy in the cage atom decays (see [23] and references therein).

The effect of Auger electron scattering leads to additional $|F_{l'}(k)|^2$ factor in the Auger width $\Gamma_{i \to f_1 f_2}^{A(A)}$ of the atom A leading to modified width in the endohedral $A@C_N$:

$$\Gamma_{i \to f_1 f_2}^{AC(A)} = F_{l'}^2(k_{i \to f_1 f_2}) \Gamma_{i \to f_1 f_2}^{A(A)}, \tag{24}$$

where $i \to f_1 f_2$ is the considered Auger-transition, $k_{i \to f_1 f_2}$ is the emitted electron linear, while $l'$ its angular momentum.

The virtual fullerenes shells excitation leads to a remarkably simple formula for the radiative width

$$\Gamma_{i \to f}^{AC(\gamma)} = \Gamma_{i \to f}^{A(\gamma)} \left| 1 - \frac{\alpha_C^d(\omega_{if})}{R^3} \right|^2 = S(\omega_{if}) \Gamma_{i \to f}^{A(\gamma)}, \tag{25}$$

where $\omega_{if}$ is the energy of the emitted in the decay photon.

Let us consider the Auger-decay in $A@C_N$ that becomes possible while it is energetically forbidden in the isolated atoms A. We will concentrate on dipole transitions $i \to f$, because only for them the radiative decay channel in the A atom is open. Using (17) we obtain for the Auger-width of a decay that for an isolated atom A is only radiative

$$\Gamma_{i \to f}^{AC(A)} = \Gamma_{i \to f}^{A(\gamma)} \frac{3}{8\pi} \left( \frac{c}{\omega_{if}} \right)^4 \frac{\sigma^C(\omega_{if})}{R^6}. \tag{26}$$

Here $\sigma^C(\omega_{if})$ is the total photoabsorption cross-section at photon energy, equal to the decay energy $\omega_{if}$, $\Gamma_{i \to f}^{A(\gamma)}$ is the radiative width of the atomic transition $i \to f$. As examples serve the subvalent vacancy decay in noble gases, for instance, the transition $3s^{-1} \to 3p^{-1}$ in Ar.

## 6. Parameters, necessary for calculations

Concrete calculations were performed for $C_{60}$ and $C_{240}$. The $C_{60}$ parameters in the present calculations were chosen the same as in the previous papers, e.g. in [14]: $R_1 = 6.72$ and $V_1 = 0.443$. For this radius and the thickness of $C_{60}$ shell equal to 1.9, our approach is well justified for photoelectron energies of about 2-3 atomic units. However, for completeness and understanding the tendencies, we present data for higher energies also.

On the bases of data in [25], we have the following values for $C_{240}$ [9]: $R_2 = 13.5$ and $V_2 = 0.53$.

To perform the calculations of the polarization amplitude factors $G(\omega)$ one needs to know dynamic polarizabilities of the fullerenes. The ab-initio calculations of these parameters are very difficult and we prefer to use here another approach. Indeed, using the relation between the imaginary part of the polarizability and the dipole photoabsorption cross-section $\sigma_C^d(\omega)$ - $\operatorname{Im}\alpha_C^d(\omega) = c\sigma_C^d(\omega)/4\pi\omega$, one can derive the polarizability from experimental data on photoabsorption of $C_N$ shell. Although experiments [26] do not provide absolute values of $\sigma_C^d(\omega)$, it can be reliably estimated using different normalization procedures on the basis of the sum rule: $(c/2\pi^2)\int_{I_C}^{\infty}\sigma_C^d(\omega)d\omega = N$, where $N$ is the number of collectivized electrons, 240 for $C_{60}$. The real part of polarizability is connected with the imaginary one (and with the photoabsorption cross-section) by the dispersion relation:

$$\operatorname{Re}\alpha_C^d(\omega) = \frac{c}{2\pi^2}\int_{I_C}^{\infty}\frac{\sigma_C^d(\omega')d\omega'}{\omega'^2 - \omega^2}, \qquad (27)$$

where $I_C$ is the $C_{60}$ ionization potential.

The equality $\operatorname{Im}\alpha_C^q(\omega) = c\sigma_C^q(\omega)/4\pi\omega$ and quadrupole dispersion relation similar to (27) are valid. Unfortunately, the quadrupole photoabsorption cross-section is so small that it cannot be derived experimentally.

Note that because we assume the strong inequality $R_{1,2} \gg r_A$ ($r_A$ being the atomic radius) we have derived formulas (17) and (18) that are more accurate than those obtained from the RPAE for the whole $A@C_N$ system. This is important since "one electron – one vacancy" channel that is the only taken into account in RPAE is not always dominant in the photoabsorption cross-section of the fullerene and hence in its polarizability.

There are no direct measurements of the photoabsorption cross section for $C_{240}$. This is why we used a scaling procedure inspired by the analyses of the data in [25] that is described in [9]. As a result we have $\alpha_2/R_2^3 \approx \alpha_1/2R_1^3$. Of course, this is a crude estimation, but our aim is to have only a feeling of what could come from the second shell in the fullerenes onion.

## 7. Results of calculations

Calculations are performed using expressions (2-6) with the amplitudes, given by expressions (19, 20) and phase shifts determined by two relations $\tilde{\delta}_{l\pm1} \equiv \delta_{l\pm1} + \Delta_{l\pm1} = \delta_{l\pm1}^{(A)} + \Delta\delta_{l\pm1}^R + \Delta_{l\pm1} + \eta^d$ and $\tilde{\delta}_{l,l\pm2} \equiv \delta_{l,l\pm2} + \Delta_{l,l\pm2} = \delta_{l,l\pm2}^{(A)} + \Delta\delta_{l,l\pm2}^R + \Delta_{l,l\pm2} + \eta^q$.

Let us start by presenting the results of the photoionization cross section at high enough energy, namely in the region of the Giant resonance in the 4d cross section in Xe. The results for Xe@$C_{60}$ and Xe are given in Fig.1 [27]. We see that one atomic Giant resonance have transformed into four *confinement resonances*. Since the effect of reflection of photoelectrons by fullerenes shell dies out with their energy growth, it is natural to see, that a Giant atomic resonance that decays by emission of fast electron remains unaltered by the fullerenes shell. This is illustrated in Fig.2 by the example of photoionization cross-section of 4*d* electrons in Eu and Eu@$C_{60}$.

The radiation enhancement parameter $S(\omega)$ for $C_{60}$ is quite big, as is demonstrated by Fig.3. To simplify the estimation of its influence, we marked by arrows thresholds of outer and subvalent subshells of all noble gas atoms. Substituting the $S(\omega)$ values at proper energies into (25), one can obtain the enhancement of the Auger-rate due to $C_{60}$ action. As to opening of Auger channel in case of only radiative decay in isolated atoms, the enhancement factor in (26) is as big as about five orders of magnitude.

Let us note a remarkable feature - $S(0) > 0$ that means that $C_{60}$ electron shell is not of metallic nature, since applied external static ($\omega = 0$) electric field is not entirely screened at the fullerenes center [24]. It appeared that $\alpha_C^d(0) \approx 60\alpha_c$, where $\alpha_c$ is the static polarizability of the carbon atom. As to the big maximum in $S(\omega)$ at $\omega \approx 25$ eV, it is a direct consequence of the Giant resonance in the $C_{60}$ photoabsorption cross section at the same energy.

Fig.4 demonstrates the effect of the second fullerene shell. It presents polarization amplitude factors $\tilde{G}_1(\omega)$ for $C_{60}$, $\tilde{G}_2(\omega)$ for $C_{240}$ and $\tilde{G}_{12}(\omega)$ for $C_{60}@C_{240}$. In Fig.5 photoionization cross-section of 3p electrons in Ar, Ar@$C_{60}$ and Ar@$C_{60}@C_{240}$ with account of reflection factors F and the latter with account of polarization factor $G_{12}(\omega)$ are presented. As FRPAE we denote the result obtained in RPAE frame with account of reflection for the intermediate state, as is described at the end of Section 3. By FRPAE2 we denote results of FRPAE with two-shell reflection taken into account.

We see that combined action of reflection and polarization transformed a smooth free-atom curve into a structure with powerful maxima called *Giant endohedral resonance*. It is remarkable that the cross-section reaches values up to about 1600 Mb that is by a factor 40-45 more than for isolated atoms and the corresponding sum rule of the resonance region is about 45-50, thus exciding the atomic value (6 for $np$ - subshell) by a factor of 8. The situation is similar at least for outer shells of Kr and Xe.

In Fig.6 as an example we present the dipole angular anisotropy parameter $\beta$ of 3p electrons for Ar and Ar@$C_{60}$. The role of reflection by $C_{60}$ is much smaller than in the cross-section, while the polarization does not affect $\beta$ at all. The moderate role of $C_{60}$ is seen also in the non-dipole angular anisotropy parameter $\gamma^C$, depicted in Fig.7, where additional not too big variations appear due to $C_{60}$.

In Fig.8 we present photoionization cross-sections of 5s electrons in Xe and Xe@$C_{60}$. Fullerene shell adds a strong maximum at 50 eV and a noticeable maximum at 110 eV. These maxima can be called *Interference endohedral resonances*.

In all considered cases the influence of the fullerenes shell upon the photoionization of the "caged" atoms is prominent. Rather impressive are the oscillations due to reflection of the photoelectron by the fullerenes shell, described by the factor $F_{l'}(\omega)$. The influence of $S(\omega)$ is also big enough.

## 8. Discussion and conclusion

It would be of interest to see the alteration of the photoionization cross-section if instead of $C_{60}$ and $C_{240}$ non-spherical fullerenes would be considered and the equilibrium position of the cage atom would be not at the center of the fullerenes shell.

The existence of Endohedral Giant and Interference resonances, along with confinement resonances is predicted. They result from strong fullerenes static and dynamic action upon atom, caged inside the fullerene. It was shown that the fullerenes shell modifies impressively the angular anisotropy parameters. We demonstrated that depending upon the photoelectron's speed the atomic Giant resonance is either destroyed, as in Xe@$C_{60}$ or remains almost untouched, as in Eu@$C_{60}$. These predictions deserve experimental verification.

Much has to be done from the theoretical side by eliminating the crude assumptions used in the above calculations and preserving at the same time the transparency of consideration. This would permit to use the photoionization of the caged atom really as a "lamp" that helps to see the fullerene from inside more effectively.

It is essential to have in mind that while being caged, the atoms inside can be ionized. The electrons go to the fullerenes shell that become instead of a neutral, a negatively charged surface. The surface charge requires some modification in the accounting for the reflection of the photoelectron by the fullerenes shell.

In atomic ground state and close to them no complexity, like quantum phase transition, exists. In fullerenes it is already enough electrons to open such a possibility. A "lamp" inside can help to disclose them if we acquire ability to decipher the signals from this "lamp" properly.

This direction of research seems to be promising and we are only at the very beginning of a long track.

**References**

1. K. Mitsuke, T. Mori, J. Kou, Y. Haruyama, and Y. Kubozono, *4d-->4f dipole resonance of the metal atom encapsulated in a fullerene cage*: Ce@$C_{82}$, Journal Chem. Phys., 122, 064304 –1-5, 2005.
2. A. Muller, S. Schippers, R. A. Phaneuf, M. Habibi, D. Esteves, J. C. Wang, A. L. D. Kilcoyne, A. Aguilar, S. Yang, and L. Dunsch, *Photoionization of the endohedral fullerene ions Sc3N@C80$^+$ and Ce@C82$^+$ by synchrotron radiation*, Journal of Physics: Conference Series **88**. 012038 –1-8, 2007.
3. V. R. Shaginyan, M. Ya. Amusia, and K. G. Popov, *Universal behavior of strongly correlated Fermi-systems*, Physics – Uspekhi, **50**(6), 1-34, 2007.
4. M. Ya. Amusia, *Atomic Photoeffect*, New York – London: Plenum Press, 1990.
5. G. Wendin and B. Wästberg, Many-electron effects in Ba$C_{60}$: Collective response and molecular effects in pptical conductivity and photoionization, Phys. Rev. B, **48**, 19, 764-767, 1993.
6. J.-P. Connerade, V. K. Dolmatov, S. T. Manson, *On the nature and origin of confinement resonances,* J. Phys. B: At. Mol. Opt. Phys. **33**, 2279-2285, 2000.
7. M. Ya. Amusia, A. S. Baltenkov and L. V. Chernysheva, *Giant Resonances of Endohedral atoms*, JETP Letters, **87**, 4, 230-233, 2008.
8. V. K. Dolmatov, P. Brewer, and S. T. Manson, Photoionization of atoms confined in giant single-walled and multiwalled fullerenes, Phys. Rev. A**78**, 013415 (2008)
9. M. Ya. Amusia, L. V. Chernysheva, and E. Z. Liverts, *Photoionization of atoms stuffed inside a two-shell fullerene*, Phys. Rev. A, submitted, 2009. On web 2009 arXiv:0904.1844 [pdf]
10. Amusia, Ivanov V. K., Cherepkov N. A. and Chernysheva L. V., *Interference Effects of Noble-Gas Atoms Outer s-Subshell*, Phys. Lett. A **40**, 5, 361-2, 1972
11. M. Ya. Amusia, N. A. Cherepkov, *Many-Electron Correlations in the Scattering Processes*, Case Studies in Atomic Physics, **5**, 2, 47-179, 1975.
12. A. S. Baltenkov, *Resonances in the photoionization cross section of M@$C_{60}$ endohedrals*, Phys. Lett. A **254**, 203-209, 1999.
13. M. Ya. Amusia, A. S. Baltenkov, V. K. Dolmatov, S. T. Manson, and A. Z. Msezane, *Confinement Resonances* in *Photoelectron Angular distributions from Endohedral atoms*, Phys. Rev. A **70,** 023201-1-5, 2004.

14. M. Ya. Amusia, A. S. Baltenkov and U. Becker *Strong Oscillations in the Photoionization of 5s-electrons in the Endohedral Atom* Xe@$C_{60}$, Phys. Rev. A **62**, 1, 012701-1-5, 2000.
15. M. Ya. Amusia, A. S. Baltenkov and L. V. Chernysheva, *On the photoionization of the outer electrons in noble gas endohedral atoms*, JETP (Zh. Exp. Teor. Fyz.) **134**, 2(8), 221-230, 2008.
16. M. Ya. Amusia, A. S. Baltenkov and L. V. Chernysheva, *Photoionization of subvalent electrons in noble gas endohedrals: interference of three resonances*, J. Phys. B: At. Mol. Opt. Phys., **41** 165201-1-7, 2008.
17. M. Ya. Amusia, A. S. Baltenkov and L. V. Chernysheva, *Distortion and preservation of Giant resonances in Endohedral Atoms* A@$C_{60}$, JETP Lett., **89**, 6, 322-326, 2009.
18. M. Ya. Amusia, P. U. Arifov, A. S. Baltenkov, A. A.Grinberg, and S. G. Shapiro *Calculation of Current Induced by Photon Momentum in Gaseous Ar*, Phys. Lett. A **47**, 1, p. 66-67, 1974.
19. M. Ya. Amusia, A. S. Baltenkov and L. V. Chernysheva, Z. Felfli Z, and A. Z. Msezane, Non-dipole Parameters in Angular Distributions of Electrons in Photoionization of Noble Gas Atoms, Phys. Rev. A **63**, 052506, 2001.
20. J. W. Cooper, Photoelectron-angular-distribution parameters for rare-gas subshells, Phys. Rev. A **47**, 1841-1851, 1993.
21. M. Ya. Amusia and L. V. Chernysheva, *Computation of Atomic Processes*. - Bristol – Philadelphia: "Adam Hilger" Institute of Physics Publishing. 1997.
22. M. Ya. Amusia. and A. S. Baltenkov, Effect of plasma oscillations of $C_{60}$ collectivized electrons on photoionization of endohedral noble-gas atoms, Phys. Rev. A **73**, 062723, 2006.
23. M. Ya. Amusia. and A. S. Baltenkov, *Vacancy Decay in Endohedral Atoms*, Phys. Rev. A **73**, 063206, 2006.
24. M. Ya. Amusia. and A. S. Baltenkov, On the possibility to consider fullerene shell $C_{60}$ as a conducting sphere, Phys. Lett. A **360**, 294-298, 2006.
25. J. M. Cabrera-Trujillo, J. A. Alonso, M. P. Iñiguez, M. J. López, and A. Rubio, *Theoretical study of the binding of Na clusters encapsulated in the $C_{240}$ fullerene*, Phys. Rev., B **53**, 16059-16066, 1996.
26. J. Berkowitz, *Sum rules and the photoabsorption cross sections of $C_{60}$*, J. Chem. Phys., **111**, 1446-53, 1999.
27. M. Ya. Amusia, A. S. Baltenkov, L. V. Chernysheva, Z. Felfli, and A. Z. Msezane, *Dramatic distortion of 4d Giant resonance by the Fullerenes $C_{60}$ shell*, J. Phys. B: At. Mol. Opt. Phys, **38**, L169-73, 2005.

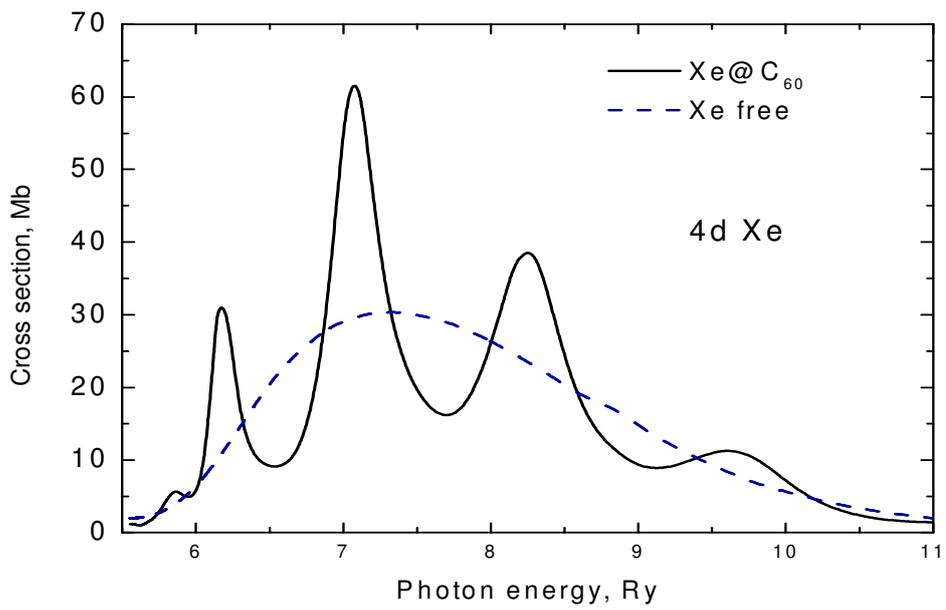

Fig.1. Photoionization cross-section of 4*d* electrons in Xe and Xe@C$_{60}$ [27].

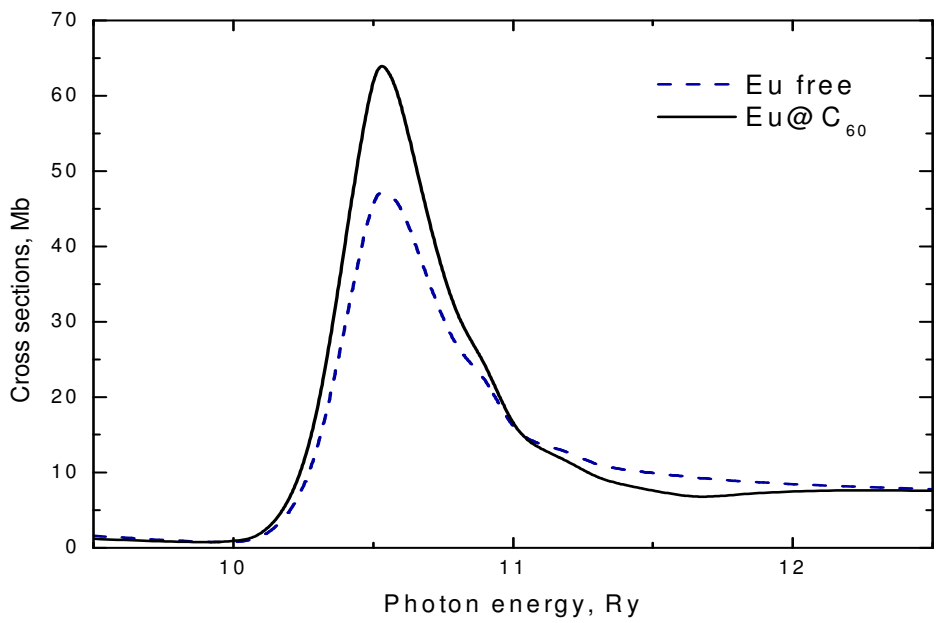

Fig.2. Photoionization cross-section of 4*d* electrons in Eu and Eu@C$_{60}$

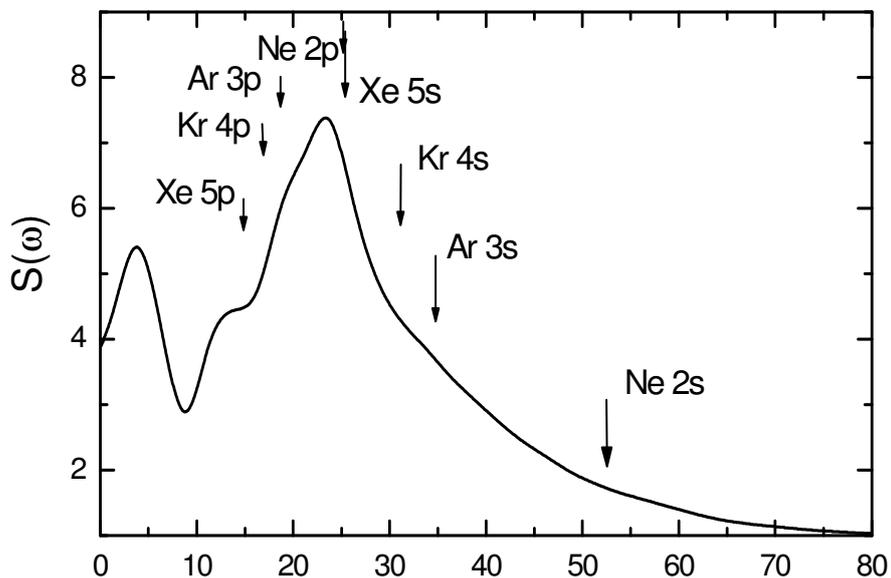

Fig.3. Radiation enhancement parameter $S(\omega)$. Arrows denote the thresholds positions of corresponding outer $np$ and subvalent $ns$ subshells.

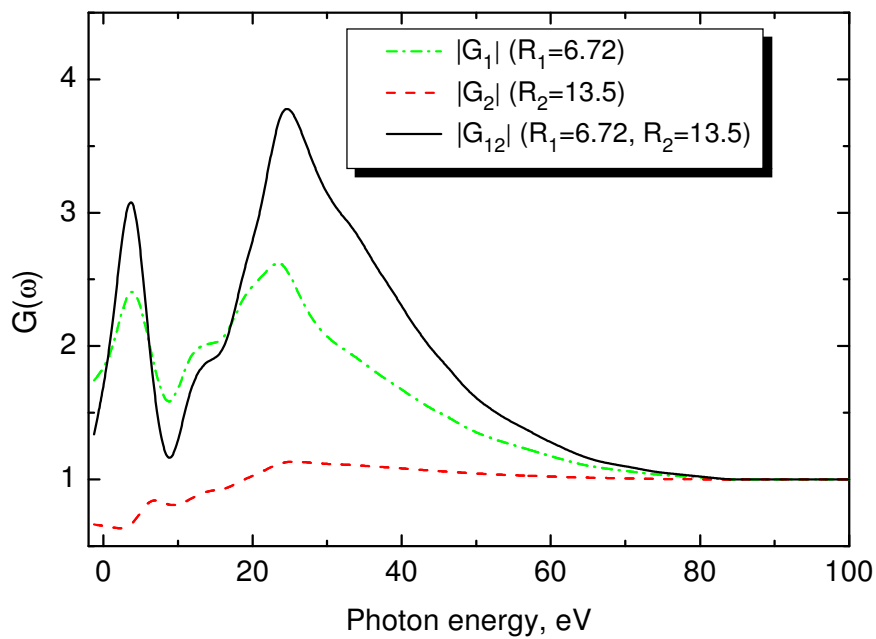

Fig.4. polarization amplitude factors $\tilde{G}_1(\omega)$ for $C_{60}$, $\tilde{G}_2(\omega)$ for $C_{240}$ and $\tilde{G}_{12}(\omega)$ for $C_{60}@C_{240}$.

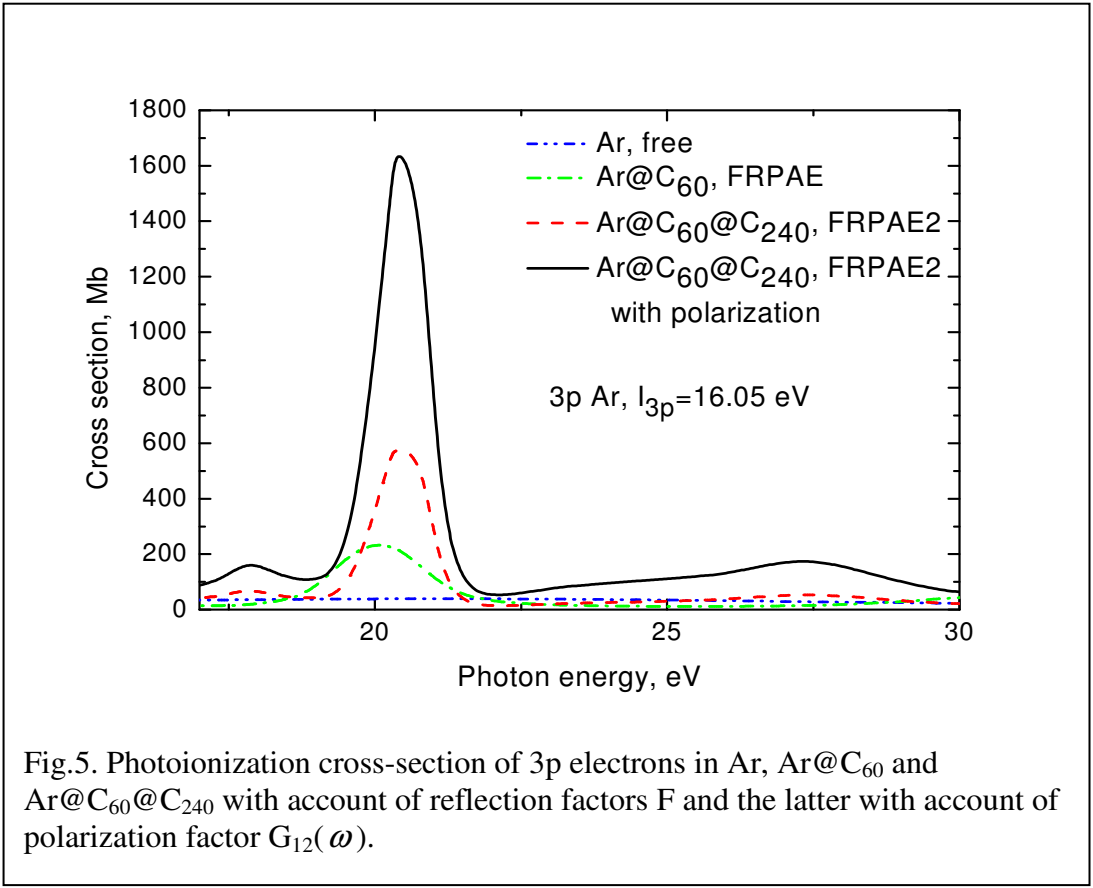

Fig.5. Photoionization cross-section of 3p electrons in Ar, Ar@$C_{60}$ and Ar@$C_{60}$@$C_{240}$ with account of reflection factors F and the latter with account of polarization factor $G_{12}(\omega)$.

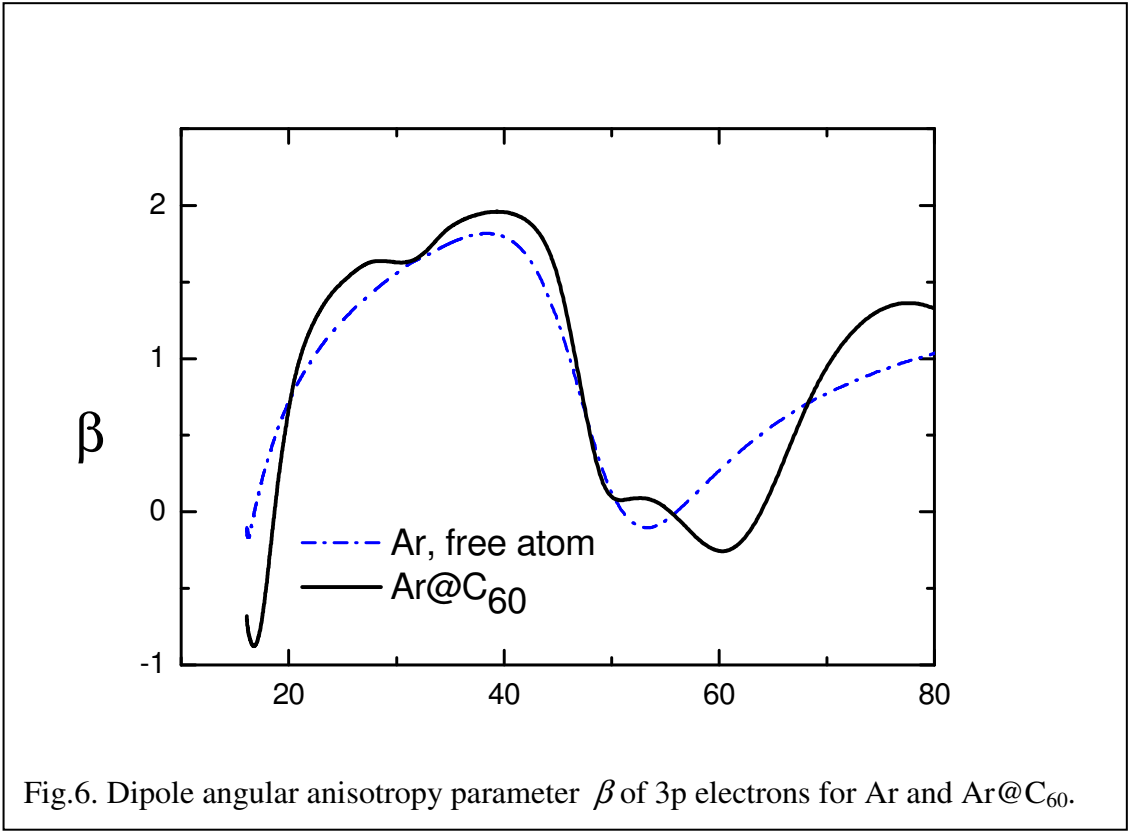

Fig.6. Dipole angular anisotropy parameter $\beta$ of 3p electrons for Ar and Ar@$C_{60}$.

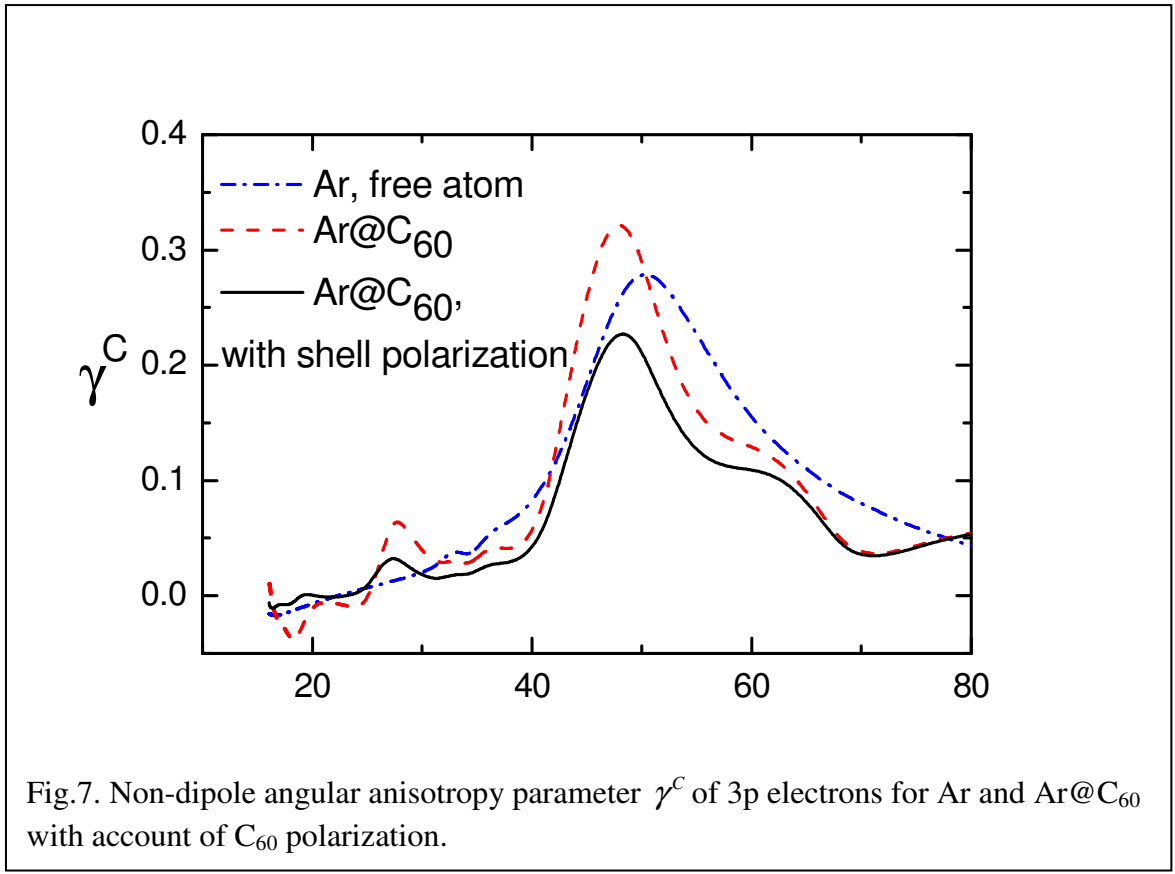

Fig.7. Non-dipole angular anisotropy parameter $\gamma^C$ of 3p electrons for Ar and Ar@$C_{60}$ with account of $C_{60}$ polarization.

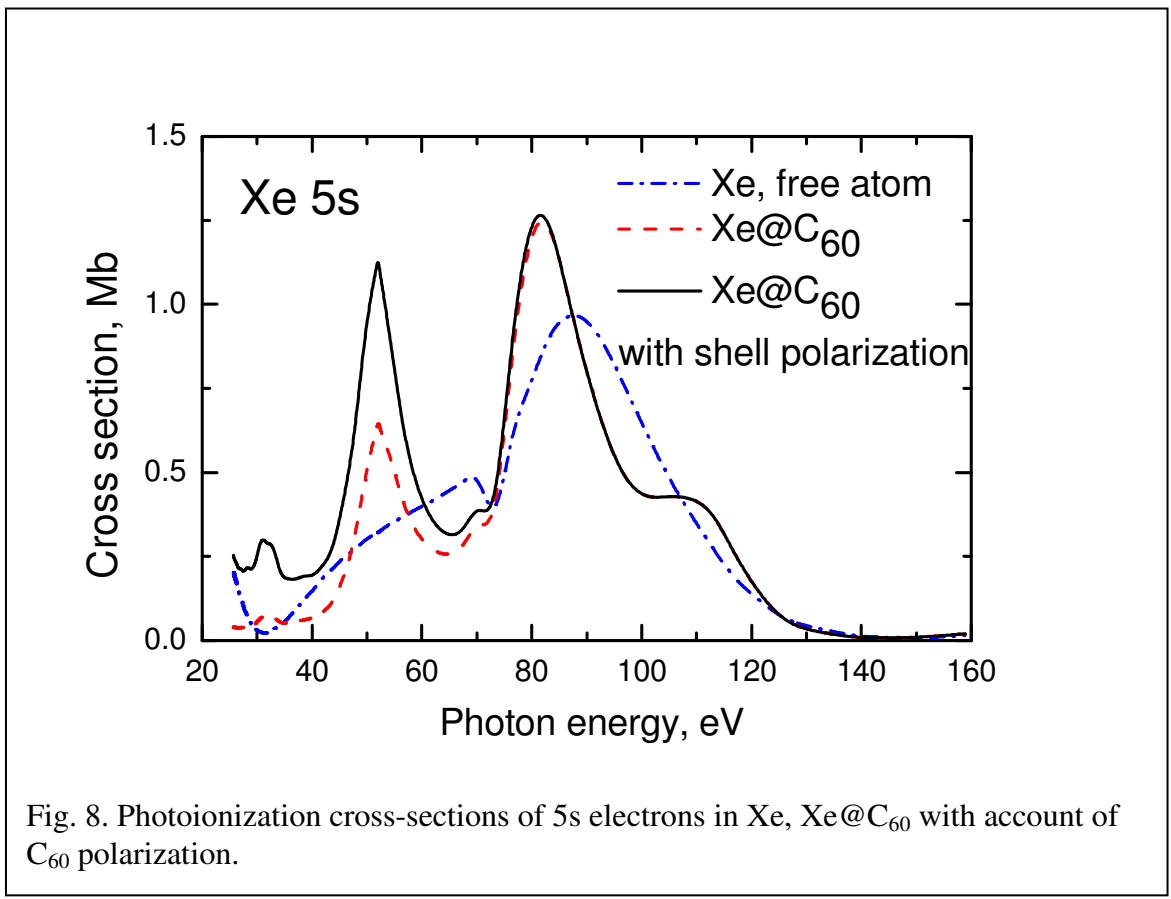

Fig. 8. Photoionization cross-sections of 5s electrons in Xe, Xe@$C_{60}$ with account of $C_{60}$ polarization.